# Pulley coupler engineering for frequency comb generation based on the supermodal approach

*Lise Morice* [1,2], *Baptiste Routier* [2,3], *Quentin Wilmart* [3], *Christian Grillet* [2], *Christelle Monat* [2], *Yoan Léger* [1]

[1]Univ Rennes, INSA Rennes, CNRS, Institut FOTON – UMR 6082, F-35000 Rennes, France

[2]Institut des Nanotechnologies de Lyon, CNRS – UMR – 5270, Ecole Centrale de Lyon, Ecully, France

[3]Univ Grenoble Alpes, CEA LETI, F38000 Grenoble, France

**ABSTRACT**

Light coupling into integrated microresonators is typically achieved using straight bus couplers that are optimized to achieve critical coupling at the pump wavelength but lack spectral engineering capabilities. Pulley couplers offer additional degrees of freedom and opportunities, but optimizing their spectral response requires intensive and time-consuming numerical simulations. To address this challenge, we developed a semi-analytical approach to quickly and efficiently calculate the coupling quality factor of pulley couplers. This enables the design of either broadband or wavelength-selective pulley couplers, depending on the target application. We illustrate the dispersion engineering capabilities offered by this new approach in the context of Kerr microcombs and experimentally validate it on the SiN photonic platform. We show that it is possible to achieve a non-dispersive broadband coupler, enabling the generation of frequency combs with similar spectral characteristics (span and intensity) irrespectively of the chosen pump wavelength.

## I. Introduction

Optical microresonators are the photonic structures of choice for generating frequency combs through 3$^{rd}$ order nonlinear phenomena due to their low mode volumes and high-quality factors, which both enhance light-matter interaction. Since the first Kerr comb demonstrated in a suspended toroidal microresonator[1], more robust designs related to integrated devices have emerged, such as microrings, which benefit from reduced manufacturing constraints, more precise dispersion engineering and simple integration with on-chip waveguides.[2–4]

The approach used for coupling light into these microrings is a critical aspect for Kerr comb generation as it affects both the conversion efficiency[5] and the threshold power[6] needed to produce the comb. The standard coupling geometry still makes use of a straight access waveguide separating by a controlled gap from the microresonator, which provides critical coupling at a specific frequency (typically that of the pump)[2,7]. Yet, this coupling strategy has several major limitations, as it generally requires a narrow gap for resonators with a high-quality factor, which might be challenging to fully control and achieve technologically, while it does not allow for any control over the coupler spectral response, since the gap is basically the only adjustable parameter. In contrast, pulley coupler geometries, where the access waveguide is wrapped around a controlled section of the microring with a constant gap, have been recently developed[8] since they offer additional degrees of freedom for optimizing the coupling performance according to the targeted application, therefore enabling the design of couplers with distinct spectral characteristics, that can be either broadband or conversely, highly selective. In the context of Kerr frequency combs, pulley couplers have thus been successfully leveraged to optimize the intensity profile of a frequency comb[8], to reduce the losses towards higher order modes[9,10] and to suppress Raman scattering interfering with frequency comb generation.[11] The interest in the dispersive nature of pulley couplers is actually not limited to the generation of frequency combs, but also offers considerable potential for a broader range of nonlinear applications, such as optical hyperparametric oscillation[12], microring modulators[13], as well as in a variety of fields including quantum optics[14].

However, the design optimization of pulley couplers is often limited to achieving phase matching between the microring and the access waveguide for a given frequency, using numerical simulations based on the Finite-Difference Time-Domain (FDTD) method in 2D or 3D[11], which are particularly computationally intensive. Although machine learning methods using neural networks have been implemented to rapidly predict the spectral response of these couplers[15], they require data-intensive training, struggle to predict “anti-resonance” where variations in coupling efficiency span several orders of magnitude, and lose accuracy outside the training range. A simple and fast calculation method capable to yield the dispersive response of this type of couplers is yet still missing in order to exploit their full potential and perform proper coupling engineering driven by the intended application.

In this work, we present a reliable semi-analytical approach to optimize the design of pulley couplers, based on supermodes of coupled waveguides and inspired by Coupled Mode Theory (CMT) in time and space. This method drastically reduces the use of resource-intensive 3D digital modeling. This new optimization approach is experimentally validated on a silicon nitride (SiN) photonic platform intended for frequency comb generation at telecom wavelengths. Following our theoretical prediction, we experimentally confirm that the geometry of a pulley coupler can be readily engineered to optimize the coupler spectral response so as to make it either highly selective or broadband. We illustrate the advantage of the latter scenario by experimentally demonstrating that frequency combs with similar span and intensity levels can be generated under a fixed pump power irrespective of the pump wavelength tuned across at least 30 nm. This coupling strategy advantageously simplifies the practical implementation of frequency comb generation, providing a flexible way to readily achieve frequency combs with a tunable coverage that is directly set by the pump wavelength, an outcome that is not possible to achieve with standard straight bus coupling strategy. More generally, our simple design approach thus opens up new opportunities for device optimization enabled by more systematic engineering of light coupling to microresonators.

## II. Theory

Coupled mode theory[16–19] (CMT) admits two distinct formulations depending on the derivation domain: a spatial approach and a temporal approach. Based on these two approaches, it is possible to determine an analytical expression for the coupling coefficient of a pulley coupler $\kappa_p$ requiring only the effective indices and propagation constants of the access guide, of the resonator, and of the supermodes of the coupled system.

Within the framework of spatial CMT, the total fields of the modes of the coupled system are expressed as a linear combination of the two non-coupled modes satisfying Maxwell's equations. This leads to a set of coupled differential equations, whose eigenvalues correspond to the propagation constants of symmetric ($\beta_{SM+}$) and antisymmetric ($\beta_{SM-}$) supermodes of the coupled system, which are represented in Figure 1.a. These eigenvalues are linked to the coupling coefficient $\kappa_S$, which is defined here as the overlap integral of the uncoupled modes, the expression of which is recalled in the Supplementary Material (eq. S1.3). The coupling coefficient is linked to the supermodes and the uncoupled modes according to:

$$\kappa_S = \sqrt{\left(\frac{\beta_{SM+} - \beta_{SM-}}{2}\right)^2 - \left(\frac{\beta_r - \beta_{wg}}{2}\right)^2} \quad (1)$$

where $\beta_r$ and $\beta_{wg}$ are the propagation constant of the isolated resonator and access waveguide respectively. The complete derivation of this expression is given in the Supplementary Material. The variation of these four propagation constants with the geometrical parameters of the system, as well as their spectral dependence can be accurately and quickly determined using 2D axisymmetric simulations based on Finite Element Method (FEM) as shown on Figure 1.a. When the modes of the resonator and the access waveguide are phase matched, strong coupling is reached, leading to the anti-crossing of the resulting supermode propagation constants, as shown in Figure 1.b for varying access waveguide width. Indeed, the phase mismatch between the resonator and the access waveguide is directly related to the curvature radius of the two coupled geometries, according to $\Delta(nR) = n_{eff,wg}R_{wg} - n_{eff,r}R_r$, where $n_{eff,wg}$, $R_{wg}$ and $n_{eff,r}$, $R_r$ are the effective index and the radius of the access waveguide and the resonator, respectively. Thus, the spectral dependence of the coupling coefficient $\kappa_S$ can be quickly inferred from that of the propagation constants of the isolated and coupled geometries.

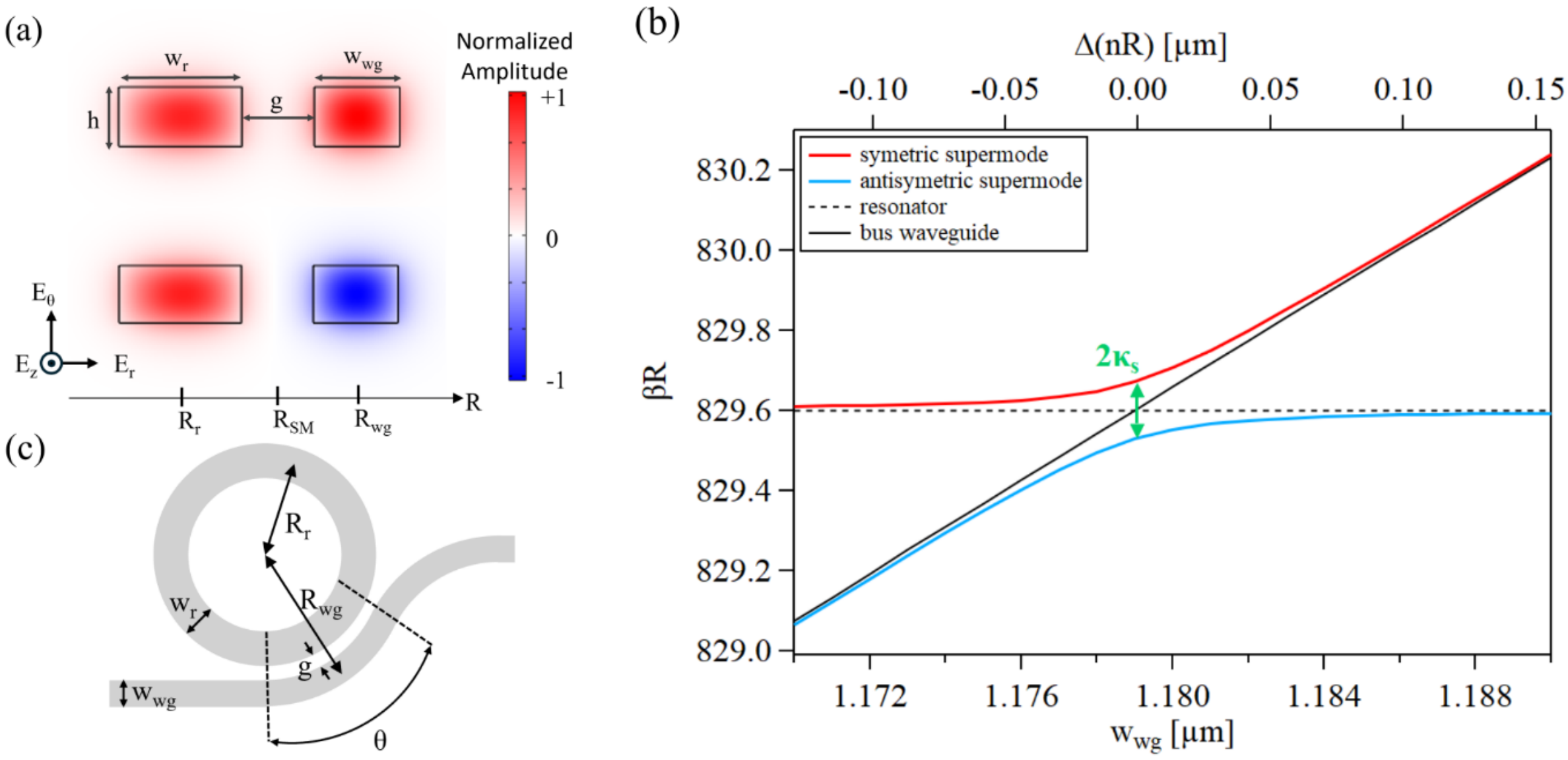

Figure 1: (a) Profile of the $E_r$ component of the electric field for the symmetric (top) and antisymmetric (bottom) $TE_{00}$ supermodes using 2D axisymmetric FEM. (b) Supermode anti-crossing curve showing the phase mismatch as a function of the access waveguide width $w_{wg}$. In the context of a 2D axisymmetric simulation, the relevant parameter is $\beta R$ rather than $\beta$, in order to take into account the geometry. (c) Schematic representation of a pulley coupler and its geometric parameters. Basically, $R_r$ and $w_r$ are fixed.

Within the framework of temporal CMT, the analysis focuses on the dynamic evolution of the field amplitude of the modes as a function of time. As derived by Manolatou *et al.*,[20] the coupling coefficient of a pulley coupler in the weak coupling approximation reads as (see Supplementary Material):

$$\kappa_p = -iS\theta e^{i\frac{\pi}{\lambda}\Delta(nR)\theta} sinc\left(\frac{\pi}{\lambda}\Delta(nR)\theta\right) \tag{2}$$

with the convention sinc(x)=sin(x)/x. Here θ is the coiling angle of the coupler, S is the overlap parameter defined such as $S = \kappa_S R_{wg}\sqrt{\frac{c}{2\pi R_r n_{g,r}}}$ with $n_{g,r}$ the group index of the resonator. All geometric parameters of the pulley coupler are represented in Figure 1.c. Note that the phase term does not appear in some articles[8], due to different choice of integration boundaries. It does not affect the physics of the coupler in any way. The coupling coefficient $\kappa_p$ is directly related to the coupling losses, so the coupling quality factor of the coupler can be defined as:

$$Q_c = \frac{2\pi c}{\lambda|\kappa_p|^2} = \frac{2\pi c}{\lambda}\frac{1}{\left[S\theta sinc\left(\frac{\pi}{\lambda}\Delta(nR)\theta\right)\right]^2} \tag{3}$$

Under this new framework, the coupling quality factor of a pulley coupler can be simply and analytically calculated from two key parameters: the field overlap integral S and the phase mismatch Δ(nR). These two factors require only the values of the effective index (or propagation constants) of the individual modes of the resonator and of the access waveguide, as well as those of the supermodes. As shown above, these parameters can be quickly and accurately determined using 2D axisymmetric numerical simulations.

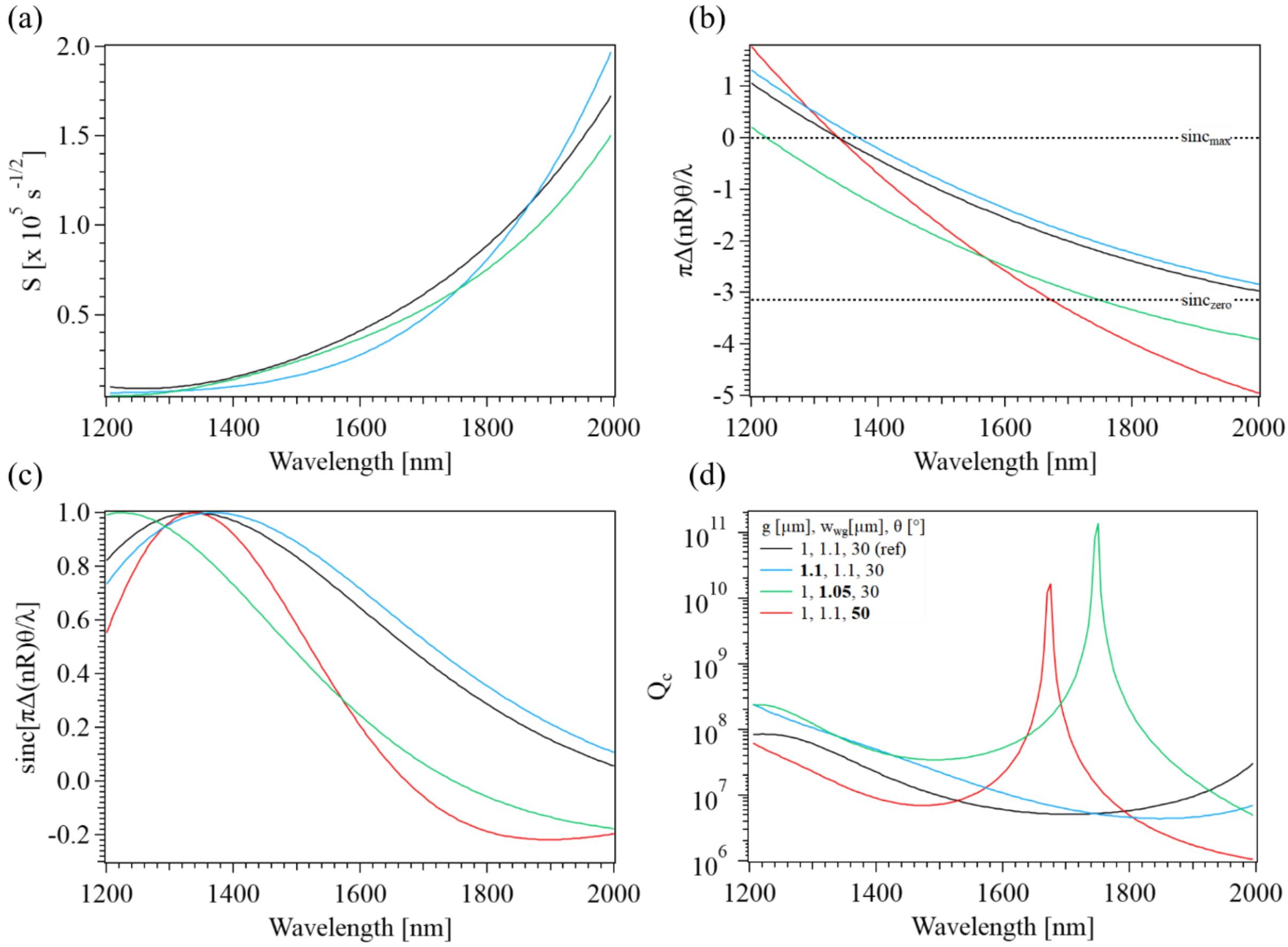

Figure 2: Impact of different geometric parameters of the pulley coupler on the coupling quality factor as a function of wavelength. Here we fixed $R_r$ = 112 µm and $w_r$ = 1.7 µm. (a) Overlap parameter S. (b) Sinc argument. (c) Sinc term. (d) Coupling quality factor. The black curve is the reference. The other curves show the effect of changing a parameter: the gap (blue), the width of the access waveguide (green) and the angle (red).

From Eq.(3), it is possible to optimize the coupling quality factor by adjusting the geometric parameters of the coupler. The gap (g) between the resonator and the access waveguide, the width of the access waveguide ($w_{wg}$) and the coiling angle (θ) are the three key geometrical parameters that can be modified for finely controlling the spectral coupling. Figure 2 shows the impact of selected coupler geometries on the spectral dependence of S, the phase mismatch, and $Q_c$. In this study, the geometrical parameters of the resonator were fixed ($R_r$ = 112 μm and $w_r$ = 1.7 μm), a relatively wide gap of 1 μm was chosen for practical fabrication reasons; the widths of the access waveguide were chosen to be close to the value ensuring phase matching at 1550 nm (i.e. 1.179 μm), and the angles were determined so as to introduce or avoid "anti-resonances" in $Q_c$ across the studied spectral range. These sets of parameters have been selected for illustration and provide a representative illustration of the intricated impact of the various parameters on the dispersive behavior of the pulley coupler, as discussed below.

First, we observe on Figure 2.a and 2.b that the gap has an influence on both the overlap parameter and the phase mismatch (comparing blue and back curves with distinct gap values). As a first approximation, the larger the gap the smaller the field overlap (i.e. the S parameter) as can be observed in Figure 2.a for wavelengths below 1870nm. For larger wavelengths and large gaps, this trend can be reversed, resulting from two joined effects: on the one hand, at high wavelengths, the modes tend to spread out spatially; on the other hand, the curvature of the access waveguide, which increases with the gap, causes a shift of the mode from the waveguide towards the resonator, thereby enhancing their interaction. In addition, a change on the gap value has an influence on the curvature radii of the access waveguide, shifting the phase mismatch conditions through $\Delta(nR)$ towards longer wavelengths, as seen in Figure 2.b. It thus also contributes to increasing the spectral position of the maximum of the sinc function (Figure 2.c). However, the phase matching condition is much more sensitive to variations in the waveguide width as shown by a comparison of the black and green curve, allowing for a precise control on the maximum and zero spectral positions of the sinc function (dotted horizontal lines of fig 2b). Thus, adjusting the position of the maximum of the cardinal sine makes it possible to either compensate for the trend of the overlap parameter caused by the effective modal section in order to obtain a spectrally flattened coupling quality factor (black curve, Figure 2.d), or to increase the slope of the overlap integral, leading to a local maximum in the coupling Q factor (green curve, Figure 2.d). Finally, θ has no direct impact on the overlap or the phase matching condition, but it allows us to control the spectral shape of the sinc argument (Figure 2.b, comparing red and black curves), and thus the spectral spread of the sinc. For a given choice of the other geometric parameters, the angle will therefore govern the presence of an "anti-resonance” (spectral band with low coupling, as on Figure 2.d, red and green curves) in the coupling quality factor over the wavelength range studied. The choice of angle can therefore allow for wavelength selectivity in the coupling efficiency, depending on the intended application.

## III. Device and quality factor

To illustrate and validate our coupling engineering approach, we implemented pulley couplers for 800 nm thick Low Pressure Chemical Vapor Deposition (LPCVD) silicon nitride ($Si_3N_4$) microrings with a 112 µm-radius used for 200 GHz frequency comb generation at telecom wavelengths[22]. The microrings are 1.7 µm-wide multimode devices encapsulated in $SiO_2$, as shown in Figure 3.a. Our analysis focused on the coupling between the fundamental $TE_{00}$ modes of the resonator and the access waveguide, without taking higher-order modes into account. Pulley couplers with a set of different geometrical parameters g, $w_{wg}$ and θ were realized with parameter settings similar to those described in the previous section, except that the angle is set to 30°. Figures 3.b and 3.c show SEM images of a pulley coupler and a close-up of the coupling area for g = 900 nm, and a wrapping angle of 30°, respectively.

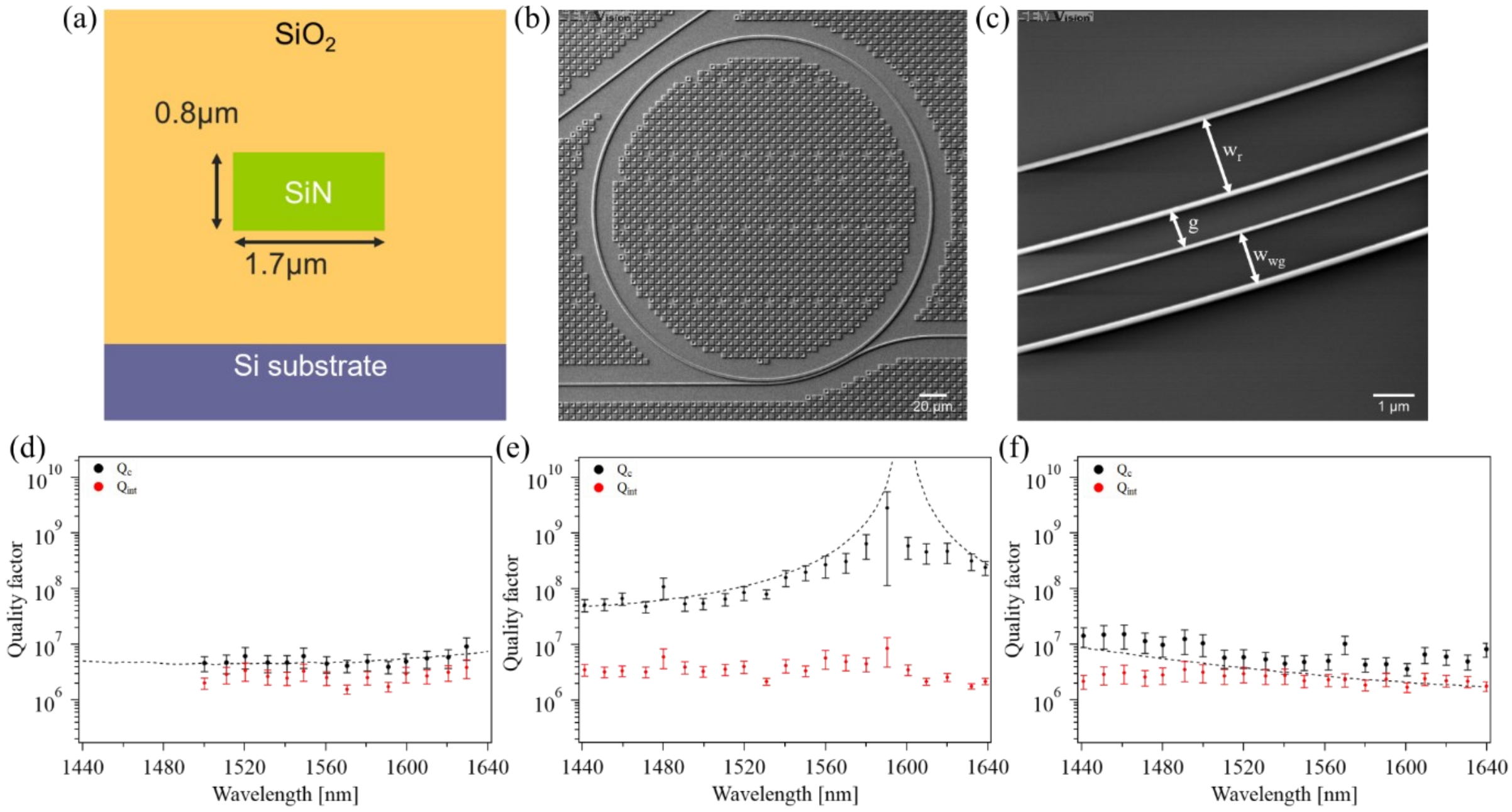

Figure 3: (a) Schematic cross-section of the SiN microring used for frequency comb at 1550 nm. (b) SEM image of a pulley coupler. (c) SEM Image of the coupling region.
Coupling quality factor (black) and intrinsic quality factor (in red) derived from the measurements as a function of wavelength for different pulley couplers with the following parameters: (d) g = 900 nm, $w_{wg}$ = 1.1 µm, θ = 30°. (e) g = 1 µm, $w_{wg}$ = 1.05 µm, θ = 30°. (f) g = 1 µm, $w_{wg}$ = 1.15 µm, θ = 30°. The dotted line corresponds to the coupling quality factor calculated using our semi-analytical method.

For each device, we measured the resonances in transmission at different wavelengths, between 1440 nm and 1640 nm, to assess the evolution of the loaded quality factor $Q_L$ and the depth of the resonance $T_0$. From these, we could infer both the intrinsic and coupling quality factors using the expression $Q_{\pm} = 2Q_L/(1 \pm \sqrt{T_0})$. The setups and measurement details are described in the Supplementary Material. Figures 3.d to 3.f show the coupling quality factors measured for different pulley couplers with slightly different sets of values for the waveguide width $w_{wg}$, gap g, and a wrapping angle fixed at 30°. We demonstrate here that the coupling quality factor trend can be profoundly modified and adjusted according to the intended use or desired performance. As discussed in section 2, depending on the parameters chosen, the sinc profile can compensate for the natural spectral dispersion of the overlap parameter S and flatten the coupler response (Figure 3.d), strengthen or counter it (see Figure 3.f and 3.e respectively), yielding or not a spectral maximum for $Q_c$.

The theoretical coupling quality factor calculated using the model described in section II is plotted as a black dashed line in Figures 3.d-f, fitting very well the experimental measurements. It should be noted here that the coiling angle $\theta$ of the pulley coupler considered in the simulation should account for the front and rear approach sections of the access waveguide where coupling cannot be neglected, especially for large radius resonators.[23] Here, these transition sections have been estimated to provide a 4° effective additional coiling angle on each side (see Supplementary Material). Furthermore, we do not observe experimentally in Figure 3.e an "anti-resonance" as narrow as that predicted theoretically. This is partially because, for an average intrinsic quality factor measured at $2.5\times10^6$ (red dots, Figure 3.d-f), coupling quality factors greater than $10^9$ result in a very shallow resonance ($T_0$ greater than 0.99), which is difficult to measure with our experimental setup. Under these low coupling conditions, higher-order modes are most likely to have been experimentally detected. Still the variation in $Q_c$ observed experimentally around the "anti-resonance" of Figure 3.e is more than one order of magnitude across a few tens of nanometers.

## IV. Effect on frequency comb generation and discussion

In this section, we focus on the effects of a flattened spectral response of the pulley coupler as shown in Figure 3.d in the context of frequency comb generation. We illustrate how it can be advantageously leveraged to produce a flexible comb response that can be spectrally tuned as per the pump wavelength, without otherwise degrading its characteristics (span, intensity). Comb generation becomes possible when the intracavity power exceeds the threshold power required for nonlinear gain to exceed losses. Above this threshold, different comb regimes can be established, such as Turing rolls, chaotic regimes, or soliton regimes. This threshold power strongly depends on the coupling and detuning conditions and is expressed as follows[24]:

$$P_{th} = \frac{\pi n_{g,r}^2 V_{eff}}{4 n_2 \lambda Q_{int}^2} \frac{(1 + Q_{int}/Q_c)^3}{Q_{int}/Q_c} + \frac{n_{g,r}^2 V_{eff} (\omega - \omega_0)^2}{2 n_2 \omega c} \left(1 + \frac{1}{Q_{int}/Q_c}\right) \quad (4)$$

where $V_{eff} = 2\pi A_{eff}$ is the effective mode volume, $n_2$ is the nonlinear refractive index of the cavity and $\omega$-$\omega_0$ is the frequency detuning of the input laser frequency from the resonance of the cold cavity. From the quality factor measurements, we can thus calculate the threshold for each resonance of a given device. In Figure 4.a, the dot markers show the coupling factor $Q_{int}/Q_c$ as a function of wavelength estimated using the measured quality factors for the pulley configuration of Figure 3.d. The solid line indicates the expected theoretical evolution of this factor using the theoretical $Q_c$ value calculated by the semi-analytical method and the average value of the intrinsic Q-factor for this material platform ($Q_{int} \approx 2.5\times10^6$). For comparison, the dotted line displays the spectral dependence of the coupling factor for a straight bus waveguide separated from the ring by a gap of 500 nm, as typically used for these microrings to generate frequency combs at 1550 nm. We see that it is possible to adjust the design of a pulley coupler to achieve a constant coupling factor across a wavelength range wider than 100 nm (from 1440 to 1580 nm), in contrast with what is achievable with the straight coupler. From the coupling factor, we can deduce the threshold power (eq. 4) as a function of wavelength for zero frequency detuning. In this type of device, the threshold power is maintained roughly constant around 10mW across the whole spectral range of the test laser, which is larger than 100 nm. Slight variations on the threshold power are induced by changes of the intrinsic quality factor from one resonance to the other (see Figure 3.d-f), a common phenomenon mostly related to local defects within the microring, but quite independent of the coupling strategy (see the Supplementary material). For pulley couplers, such low power thresholds should be preserved down to 1400 nm as shown by the theoretical solid line curve which is flatter across a broader range than the dotted line obtained for straight couplers. Additional details on the coupling factor and threshold power measurements in the straight coupler case are included in the Supplementary Material.

To demonstrate the versatility of the pulley coupler suggested by the spectrally flat response shown on Figure 4.a-b, we have generated frequency combs at different pump wavelengths ranging between 1530 and 1560 nm, and a fixed power of 50 mW as shown in Figure 4.c. The optical setup used to generate the frequency combs is the same as that used for quality factor measurements presented in the Supplementary Material, except that the pump wavelength is fixed and the laser signal amplified with an Erbium-Doped Fiber Amplifier. Figure 4.c shows that the spectral signature and span of the four obtained combs are very similar, irrespective of the pumped resonance, thereby giving some evidence of the relatively constant coupling regime. In contrast, for the straight coupler, as the coupling is not the same at the different pump wavelengths, the generated combs have strikingly distinct spectral signatures, as shown in

Figure 4.d. The use of pulley couplers will be critically important when moving towards the co-integration of the laser source on photonic chips, as this will allow for greater flexibility in terms of the laser's emission wavelength enabling the generation of a comb.

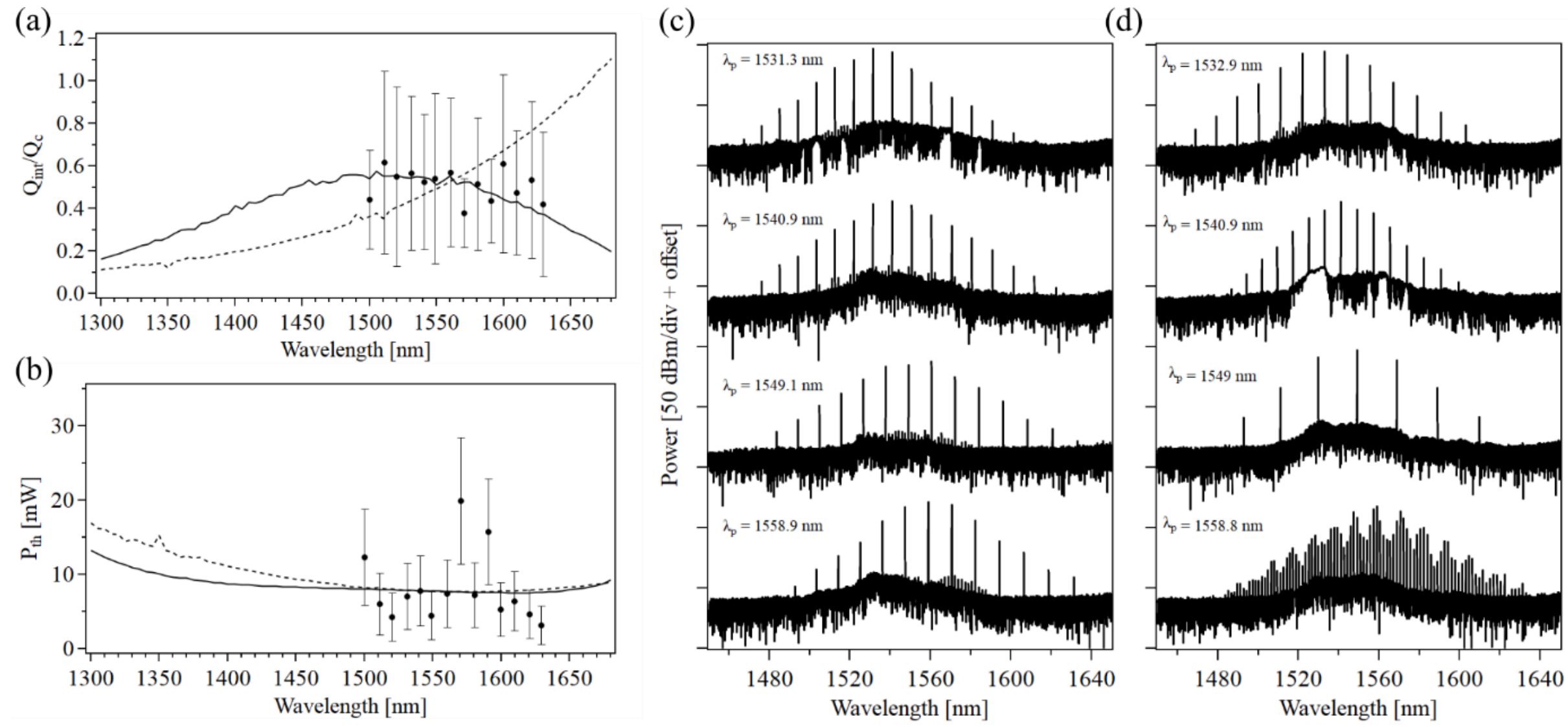

Figure 4: (a) Coupling factor as a function of wavelength with zero frequency detuning for a pulley coupler with geometrical parameters g = 900 nm, $w_{wg}$ = 1.1 µm and θ = 30°. The dots correspond to the experimental Q-factor measurements and the solid line to the theoretical calculations using the semi-analytical approach. The dotted line corresponds to the theoretical coupling factor for a straight coupler with a gap of 500 nm. (b) Threshold power as a function of wavelength for the same coupler. Frequency comb generation in a 112 µm-radius ring resonator coupling with (c) a pulley coupler (g = 900 nm, $w_{wg}$ = 1.1 µm and θ = 30°) and (d) a straight coupler with g = 500 nm. From top to bottom, the pump wavelength increases, and the pump power is fixed around 50 mW.

## V. Conclusion

In conclusion, we have presented a semi-analytical approach for the design of pulley couplers, based on supermodes computation. The analysis of the trends yielded by this new, reliable and simple design method clarifies the interplay between the multiple parameters of the pulley coupler to control its spectral response, while proving its key advantages for device optimization. Our study highlights the ability to profoundly control the coupler spectral response via adjusting its geometry, enabling either constant or highly selective coupling, which could be alternatively preferred depending on the intended application. The possibility to reach these distinct regimes were experimentally validated from measurements conducted on the mature SiN material platform, where good agreement with our semi-analytical model was found. We have illustrated the potential interest afforded by designing pulley couplers with a roughly constant coupling ratio (across more than 100nm) in the context of frequency comb generation, where a comb spectrum with similar features (relative intensity and span) could be generated from a single device, irrespective of the pump wavelength. Conversely, for a fixed pump power, similar frequency combs could be observed in a flexible way, only detuned in wavelength by more than 30nm, as set by the chosen pump wavelength. This flexibility to operate the comb could prove particularly useful for the co-integration of a Kerr-comb resonator with a pump laser yielding a limited tunability range. More generally, our new design approach provides an effective and useful method to quickly optimize and engineer the dispersive coupler response according to the targeted application.

**Supplementary Material**

The Supplementary Material presents the CMT, applied to spatial and temporal approaches, enabling the derivation of an expression for the coupling coefficient of a pulley coupler. It also explains the optical setup used to measure the fine resonances of the resonator and details for calculating the different quality factors. There is also a comparison with a straight coupler in the context of frequency comb generation.

**Acknowledgement**

The authors acknowledge nanoRennes for the technological support, a platform affiliated to RENATECH+ (the French national facilities network for micro-nanotechnology). This research was supported by "France 2030" with the French National Research agency OFCOC project (ANR-22-PEEL-0005).

**Author Declarations**

**Conflict of interest**

The authors have no conflicts to disclose.

**Author Contributions**

**Lise Morice :** Conceptualization (equal); Investigation (lead); Validation (lead); Writing – original draft (lead). **Baptiste Routier :** Conceptualization (equal); Resources (equal). **Quentin Wilmart :** Conceptualization (equal); Resources (equal); Supervision (equal). **Christian Grillet** : Conceptualization (equal); Supervision (equal). **Christelle Monat :** Conceptualization (equal); Supervision (equal). **Yoan Léger :** Conceptualization (equal); Supervision (equal).

**Data Availability**

The data that supports the findings of this study are available from the corresponding authors upon reasonable request.